\documentclass[prd,a4paper,nofootinbib,showpacs,superscriptaddress]{revtex4}
\usepackage{bm}
\usepackage{amsfonts}
\usepackage{graphics}
\begin{document}
%-------------------------------------------------------------------------
% Definition needed for the heading
%-------------------------------------------------------------------------
\def\Barcelo{Barcel\'o}
\def\Schrodinger{Schr\"odinger}
%-------------------------------------------------------------------------
\title[On the thin-shell limit ...]
{On the thin-shell limit of branes in the presence of Gauss-Bonnet 
interactions}
%-------------------------------------------------------------------------
\author{Carlos \Barcelo}
%\email{carlos.barcelo@port.ac.uk}
%\homepage{}
%\thanks{Supported by the EC under contract HPMF-CT-2001-01203}
%\altaffiliation{}
\affiliation{Institute of Cosmology and Gravitation,
University of Portsmouth, Portsmouth PO1 2EG, Britain.}
%-------------------------------------------------------------------------
\author{Cristiano Germani}
%\email{cristiano.germani@port.ac.uk}
%\homepage{}
%\thanks{}
%\altaffiliation{}
\affiliation{Institute of Cosmology and Gravitation,
University of Portsmouth, Portsmouth PO1 2EG, Britain.}
%\affiliation{Universit\'a ``La Sapienza'' di Roma - Dipartimento di Ingegneria
%Aereospaziale e Astronautica - via Eudossiana 16, 00184 Roma, Italy.}
%-------------------------------------------------------------------------
\author{Carlos F. Sopuerta}
%\email{carlos.sopuerta@port.ac.uk}
%\homepage{}
%\thanks{}
%\altaffiliation{}
\affiliation{Institute of Cosmology and Gravitation,
University of Portsmouth, Portsmouth PO1 2EG, Britain.}
%-------------------------------------------------------------------------

%-------------------------------------------------------------------------
%\date{19 February 2003; gr-qc/0306072; \LaTeX-ed \today}
%-------------------------------------------------------------------------
\begin{abstract}
%-------------------------------------------------------------------------

In this paper we study thick-shell braneworld models in the presence
of a Gauss-Bonnet term. We discuss the peculiarities of the attainment
of the thin-shell limit in this case and compare them with the same
situation in Einstein gravity. We describe the two simplest families
of thick-brane models (parametrized by the shell thickness) one can
think of. In the thin-shell limit, one family is characterized by the
constancy of its internal density profile (a simple structure for the
matter sector) and the other by the constancy of its internal
curvature scalar (a simple structure for the geometric sector). We
find that these two families are actually equivalent in Einstein
gravity and that the presence of the Gauss-Bonnet term breaks this
equivalence. In the second case, a shell will always keep some
non-trivial internal structure, either on the matter or on the
geometric sectors, even in the thin-shell limit.

%-------------------------------------------------------------------------
\end{abstract}
%-------------------------------------------------------------------------
%-------------------------------------------------------------------------
\maketitle
%-------------------------------------------------------------------------
%-------------------------------------------------------------------------
% Author-specific definitions
%-------------------------------------------------------------------------
\newcommand{\mb}[1]{\mbox{\boldmath $#1$}}
\def\half{{1\over 2}}
\def\L{{\mathcal L}}
\def\S{{\mathcal S}}
\def\d{{\mathrm{d}}}
\def\x{{\mathbf x}}
\def\v{{\mathbf v}}
\def\im{{\rm i}}
\def\etal{{\emph{et al\/}}}
\def\det{{\mathrm{det}}}
\def\tr{{\mathrm{tr}}}
\def\ie{{\emph{i.e.}}}
\def\bnabla{\mbox{\boldmath$\nabla$}}
\def\Box{\kern0.5pt{\lower0.1pt\vbox{\hrule height.5pt width 6.8pt
    \hbox{\vrule width.5pt height6pt \kern6pt \vrule width.3pt}
    \hrule height.3pt width 6.8pt} }\kern1.5pt}
\def\HRULE{{\bigskip\hrule\bigskip}}
%-------------------------------------------------------------------------
\newcommand{\eq}{a^4(\Phi+\alpha\Phi^2+\frac{1}{\ell^2})}
\newcommand{\K}{K_{rr}}

%---------------------------------------------------------------------
\section{Introduction}
\label{S:introduccion}
%---------------------------------------------------------------------

In the light of the Randall-Sundrum (RS) braneworld paradigm \cite{RS}, the
cosmological evolution of the universe would acquire important
modifications at early times with respect to the standard lore
provided by the Friedmann equation of General Relativity
\cite{Binetruy,Friedmann,Bruck} (an exhaustive study of the phase
space of the new cosmological models can be found in~\cite{CamSop}).
The new Friedmann-like equation has now contributions that are
quadratic in the density, dark-radiation contributions and possibly
even other types of dark contributions (coming for example from a
fundamental electromagnetic field in the 
bulk \cite{Barcelo-edge,Biswas,Gregory}).

This same paradigm, with the characteristic presence 
of extra dimensions, naturally suggests that we should look for 
further modifications to the generalized Friedmann 
equation caused by the existence of a Gauss-Bonnet (GB) 
term in the field equations. In five dimensions, the most 
general geometric field equations of 
the form ${\cal G}_{AB}=0$, with ${\cal G}_{AB}$ a symmetric 
and conserved tensor constructed by using the metric 
and its first and second derivatives only, is a linear combination
of the metric tensor itself, the Einstein tensor and the Lanczos
tensor (or GB term)~\cite{Lanczos,Lovelock}. From this point
of view, the value of the three coefficients of this combination, 
basically, the cosmological constant, Newton's constant and 
the $\alpha$ constant (to be defined later) respectively, 
are parameters to be determined experimentally\footnote{We
are thinking on the standard RS scenario, in which there is
only one additional extra dimension; for 7 or more extra
dimensions one will have to consider more general Lovelock
terms~\cite{Lovelock}.}.
The GB term has particular relevance when
considering the string-inspired nature of the RS proposal.
In string theory, the GB term appears as the first higher
curvature correction to Einstein gravity
\cite{Zwiebach,Boulware,Candelas,Zumino}.
Possible consequences of the GB interactions
on cosmological inflation have already been considered
in~\cite{Lidsey,Lidsey2}. Other analysis concerning
different aspects of GB corrections to braneworld physics
can be found in \cite{Deruelle-Dolezel}--\cite{Lidsey-Nojiri}
(this is not intended to be an exhaustive list).

In trying to generalize the thin-shell cosmological models to incorporate
the effects of a GB term it has been some controversy. First, it was
claimed that the GB equations were not well defined in the
distributional sense required by the thin-shell formalism
\cite{Deruelle-Dolezel}. It would be necessary to study
the microphysics to solve the ambiguities that would arise.
However, later, in references \cite{Charmousis}--\cite{Abdesselam:2001ff},
it was assumed that the
particular structure of the GB Lagrangian (linear in
second order derivatives) posed no problems in the distributional
limit (at least in models with $\mathbb{Z}_2$ symmetry).  Remarkably,
the results obtained in those paper were not consistent with each
other. The generalized Friedmann equation found in papers
\cite{Germani}--\cite{Abdesselam:2001ff}
(quadratic on the brane density $\rho_b$) is
different from that found in papers
\cite{Charmousis}--\cite{Binetruy:2002ck}
(with a complicated dependence on $\rho_b$
coming from obtaining the real root of a cubic equation).
More recently, there has been an argument \cite{Davis,Gravanis}
based on the form of the surface term in GB (see \cite{Myers})
that strongly support this later complicated
$\rho_b$ dependence (see also \cite{Deruelle-Madore}).

In this paper, we will study the cosmological behaviour of shells
(or branes) that are thin but still of a finite
thickness $T$. In this way we want to shed some light on how
the zero thickness is attained in the presence of GB interactions.
This limit has been studied for Einstein gravity in \cite{Mounaix}.
Thick shells in the context of GB interactions
have been already studied in \cite{Corradini} and \cite{Giovannini}, but
with a focus on different aspects than those in this paper.
The conclusion of our analysis here is twofold.
On the one side, our results show that the generalized Friedmann equation
in~\cite{Charmousis}--\cite{Binetruy:2002ck} can be found by using
a completely general procedure, in which  the energy density of the brane
in the thin-limit is related to the averaged density.  Instead, the equation
in~\cite{Germani}--\cite{Abdesselam:2001ff} can only be found for
specific geometric configurations and with a procedure in which
the energy density of the brane in the thin-limit comes from
the value of the boundary density in the thick-brane model.
On the other side, we argue that the information lost when treating a
real thin shell
as infinitely thin is in a sense larger in Einstein-Gauss-Bonnet
gravity than in the analogous situation in standard General Relativity.

Let us explain further this last point. From a physical point of view,
in the process of passing from the notion of function to that of
distribution one loses information. Many different series of functions
define the same limiting distribution. For example, the series
\begin{equation}
f_T(y)=\left\{
\begin{array}[c]{l}
0 ~~~{\rm for}~~~ |y| > T/2\,, \\
\mbox{} \\
{1 \over T}  ~~~{\rm for}~~~ |y| < T/2\,;
\end{array}\right.
~~~~~~~~
g_T(y)=\left\{
\begin{array}[c]{l}
0 ~~~{\rm for}~~~ |y| > T/2\,, \\
\mbox{} \\
{12 y^2 \over T^3} ~~~{\rm for}~~~ |y| < T/2\,,
\end{array}\right.
\end{equation}
define the same limiting Dirac's delta distribution. The distribution
only takes into account the total conserved area delimited by the
series of functions. The gravitational field equations relate geometry
with matter content.  If we take the matter content to have some
distributional character, the geometry will acquire also a
distributional character. When analyzing the thin-limit of branes in
Einstein gravity, by constructing series or families of solutions
parametrized by their thickness, we observe that the blowing up parts
of the series of functions that describe the density-of-matter profile
transfer directly to the same kind of blowing up parts in the
description of its associated geometry. Very simple density profiles
[like the previous function $f_T(y)$] are associated with very simple
geometric profiles, and vice-versa.  However, when considering
Einstein-Gauss-Bonnet gravity this does not happen. The blowing up parts
of the series describing the matter density and the geometry are
inequivalent. A simple density profile does not correspond to a very
simple geometric profile and vice-versa; on the contrary, we observe
that they have some sort of complementary behaviour. This result
leads us to argue that the distributional description of the
cosmological evolution of a brane in Einstein-Gauss-Bonnet gravity
is hiding important aspects of the microphysics, not present when
dealing with pure Einstein gravity. Also, we find that for simple models
of the geometry, one can make compatible the two seemingly distinct
generalized Friedmann equations found in the literature.

The paper is organized as follows. In the next Section we build
thick-shell models for static branes (direct generalization of the RS
model \cite{RS}). This provides a simple situation in which the main
ideas of the paper can already be seen. For clarity, we will separate
the Einstein and Einstein-Gauss-Bonnet cases.  Section
\ref{S:dynamic} will deal with the dynamical cosmological case.
Finally, we will make a brief summary of the results found in Section
\ref{S:summary}.  General formulae for the construction of the field
equations are given in the Appendix.

%---------------------------------------------------------------------
\section{Static thick shells in Einstein and Gauss-Bonnet}
\label{S:static}
%---------------------------------------------------------------------

%---------------------------------------------------------------------
\subsection{Einstein}
%---------------------------------------------------------------------

To fix ideas and notation let us first describe the simple case
in which we have a static thick brane in an anti-de Sitter (adS)
bulk. We take an ansatz for the metric of the form\footnote{Five-, four-,
and three-dimensional indices are written using upper-case Latin, Greek,
and lower-case Latin letters respectively.}
\begin{eqnarray}
ds^2=  e^{-2A(y)} \eta_{\mu\nu}dx^\mu dx^\nu +dy^2\,,
\label{brane}
\end{eqnarray}
where $\eta_{\mu\nu}$ is the four-dimensional Minkowski metric.
Comparing with the formulas given in Appendix~\ref{appa} [in particular
equation~(\ref{themetric})] this means
taking $a(t,y)=n(t,y)=\exp(-2A(y))$, and $b(t,y)=1$.
The energy-momentum tensor has the form
\begin{eqnarray}
\kappa^2_5\, T_{AB} = \rho u_{A}u_{B} + p_Lh_{AB} + p_Tn_{A}n_{B}\,,
\label{tAB}
\end{eqnarray}
where $u_A=(-e^{2A},\mb{0},0)$, $n_A=(0,\mb{0},1)$
and $h_{AB}=g_{AB} + u_A u_B - n_A n_B$ with $g_{AB}$ the 
5-dimensional bulk metric. Here, $\kappa^2_5$ denotes the 
bulk gravitational coupling constant and $\rho$, $p_L$
and $p_T$ represent respectively the energy density, the 
longitudinal pressure and the transverse pressure, 
and are taken to depend only on $y$.
The Einstein equations $G_{AB}=-\Lambda_5 g_{AB}+\kappa_5^2 T_{AB}$
with a negative cosmological constant,
$\Lambda_5\equiv -6/\ell^2$,
result on the following independent equations for the metric function
$A(y)$:
\begin{eqnarray}
&&3A''-6A'^2=\rho-{6 \over \ell^2}\,, \label{eins1}  \\
&&6A'^2=p_T+{6 \over \ell^2}\,, \label{eins2} \\
&&p_L=-\rho\,.
\label{E-equations}
\end{eqnarray}
For convenience, we will hide the $\kappa_5^2$ dependence inside the matter magnitudes,
$\rho=\kappa_5^2\rho_{\rm true}$, etc.
We also consider a $\mathbb{Z}_2$-symmetric geometry around $y=0$.
The brane extend in thickness from $y=-T/2$ to $y=+T/2$. Outside
this region $\rho=p_T=0$, so we have a purely adS spacetime:
$A(y)=-y /\ell+b$ for $y\in (-\infty,-T/2)$ and $A(y)=y /\ell+b$ for
$y\in (T/2,+\infty)$ where $b$ is an irrelevant constant. 
The junction conditions at $y=-T/2,+T/2$ 
[see Eqs.~ (\ref{cindm}) and (\ref{cincm})  
tell us that
\begin{eqnarray}
&& A(-T^-/2)=A(-T^+/2),~~~~ A(T^-/2)=A(T^+/2) \,, \\
&& A'(-T^-/2)=A'(-T^+/2),~~~~ A'(T^-/2)=A'(T^+/2)\,.
\label{E-junctions}
\end{eqnarray}
{}From here and using (\ref{eins2}) we deduce that the
transversal pressure is zero at the brane boundaries
$p_T(-T/2)=p_T(T/2)=0$. Since we are imposing a
$\mathbb{Z}_2$-symmetry with $y=0$ as {\em fixed point}, hereafter we
will only specify the value of the different functions in the interval
$(-T/2,0)$.

The function $A'$ is odd and therefore interpolates
from $A'(-T/2)=-1/\ell$ to $A'(0)=0$. If in addition
we ask for the null-energy condition $\rho+p_T=3A'' \geq 0$ to be satisfied
everywhere inside the brane, then $p_T$ has to be a negative and monotonically
decreasing function from $p_T(-T/2)=0$ to $p_T(0)=-6/\ell^2$.
This condition will turn to be fundamental in defining
a thin-shell limit.

By isolating $A''$ from equations (\ref{eins1}) and (\ref{eins2}) 
we can relate the
total bending of the geometry on passing through the brane
with its total $\rho+p_T$
\begin{eqnarray}
{6 \over \ell} =3A'|_{-T/2}^{T/2}=\int_{-T/2}^{T/2}  (\rho+p_T)\; dy.
\end{eqnarray}
At this stage of generality,
one can create different one-parameter families of thick-brane versions
of the Randall-Sundrum thin brane geometry, by parameterizing each
member of a given family by its thickness $T$. The only requirement needed
to do this is that the value of the previous integral must be kept fixed independently
of the thickness of the particular thick-brane geometry.   Thus, each particular
family can be seen as a regularization of Dirac's delta distribution.

We can realize that, provided the condition $\rho+p_T \geq 0$ is satisfied,
there exists a constant $C$, independent of the thickness $T$, such that
$p_T<C$, that is, the profile for $p_T$ is bounded and will not
blow up in the thin-shell limit.
Therefore, in the limit in which the thickness of the branes goes to
zero, $T\rightarrow 0$, the integral of $p_T$ goes to zero with the
thickness. (Strictly speaking, the thin-shell limit is reached when
$T/\ell\rightarrow 0$ but throughout this paper we are going to
maintain $\ell$ constant.)  Instead, the profile of $\rho$ has to
develop arbitrarily large values in order to fulfil
\begin{eqnarray}
{6 \over \ell} =\lim_{T\rightarrow 0} \int_{-T/2}^{T/2}  \rho\;dy.
\label{average}
\end{eqnarray}

In the thin-shell limit, we can think of Einstein's equations as
providing a relation between the characteristics of the density
profile and the shape of the internal geometry. A very complicated
density profile will have associated a very complicated function
$A(y)$.  Physically we can argue that when a shell becomes very thin
one does not care about its internal structure and, therefore, one
tries to describe it in the most simple terms. But what it is exactly
the meaning of simple? Here we will adopt two different definitions of
simple: The first one is to consider that the internal density is
distributed homogeneously throughout the shell when the shell becomes
very thin. The second case is to consider that the profile for $A'$ is
such that it interpolates from $A'(-T/2)=-1/\ell$ to $A'(0)=0$ through
a straight line, or what is the same, that the internal profile of
$A''$ is constant. Again, we require this for very thin shells. This
geometric prescription is equivalent to asking for a constant internal
scalar curvature, since $R=8A''-20A'^2$ and for every thin shell the term
$A'^2$ is negligible with respect to the constant $A''$
term. Hereafter, we will use indistinctly the names {\em straight
interpolation} or {\em constant curvature} for these models. In
building arbitrarily thin braneworld models, one needs that the profiles
for the internal density $\rho$ and the internal $A''$ acquire
arbitrarily high values (they will become distributions in the limit of
strictly zero thickness). In the first of the two simple models
described, the simplicity applies to the blowing up parts of the
matter content; in the second, the simplicity applies to the blowing
up parts of the geometry. 
{}From the physical point of view advocated in the introduction, these 
simple profiles are those that do not involve losing information in the 
process of taking the limit of strictly zero thickness.

Let us analyze both cases independently.

%---------------------------------------------------------------------
\subsubsection{Constant density profile}
%---------------------------------------------------------------------

Let us first define for convenience $z\equiv y/T$ as a scale invariant
coordinate inside the brane. Then,
mathematically, the idea that the density profile, which we will assume
to be analytic inside the brane for simplicity, becomes constant
in the thin-shell limit can be expressed as follows:
\begin{eqnarray}
\rho(z)=\sum_n \beta_n(T) \; z^{2n},
\label{constant-density-est}
\end{eqnarray}
where
\begin{eqnarray}
\lim_{T\rightarrow 0} T ~ \beta_n(T)\rightarrow 0,
~~~~\forall n\neq 0;~~~~
\lim_{T\rightarrow 0} T ~ \beta_0(T)\rightarrow \rho_b : {\rm constant}\,.
\label{condi}
\end{eqnarray}
For these density profiles, the Einstein equations in the
thin-shell limit tell us that
\begin{eqnarray}
3A''=\beta_0(T)-{6 \over \ell^2}+6A'^2\,.
\end{eqnarray}
{}From here we get the profile for $A'$:
\begin{eqnarray}
A'=\sqrt{{\beta_0(T) \over 6}-{1 \over \ell^2}}
\tan \left(2\sqrt{{\beta_0(T) \over 6}-{1 \over \ell^2}}\; y \right)\, .
\label{apri}
\end{eqnarray}
Notice that this expression only makes sense for $\beta_0(T)>6/\ell^2$,
but this is just the regime we are interested in. We have to impose now
the boundary condition $A'(T/2)=1/l$ to the previous expression (\ref{apri})
\begin{eqnarray}
{1 \over \ell}=\sqrt{{\beta_0(T) \over 6}-{1 \over \ell^2}}
\tan \left(\sqrt{{\beta_0(T) \over 6}-{1 \over \ell^2}}\; T \right).
\end{eqnarray}
In this manner, we have implicitly determined the form of the function
$\beta_0(T)$. In the limit
in which $T\rightarrow 0$ with $T\beta_0(T)\rightarrow \rho_b$,
we find the following relation
\begin{eqnarray}
{6 \over \ell}=\rho_b.
\end{eqnarray}
This condition is just what we expected from the average 
condition (\ref{average}).

%---------------------------------------------------------------------
\subsubsection{Straight interpolation}
%---------------------------------------------------------------------

In this case, the mathematical idea that in thin-shell limit 
the profile for $A'$ corresponds to a straight interpolation,
can be formulated as
\begin{eqnarray}
A''(z)=\sum_n \gamma_n(T) \; z^{2n}, 
\label{straight-interpolation}
\end{eqnarray}
with
\begin{eqnarray}
\lim_{T\rightarrow 0} T ~ \gamma_n(T)\rightarrow 0,
~~~~\forall n\neq 0;~~~~
\lim_{T\rightarrow 0} T ~ \gamma_0(T)\rightarrow {2 \over \ell},
\label{straight-interpolation-cond}
\end{eqnarray}
For these geometries we find that the associated profiles for
$p_T$ and $\rho$ in the thin-shell limit have the following form
\begin{eqnarray}
&&p_T=-{6 \over \ell^2}\left(1 - 4 z^2 \right) + \omega_1(T,z) \,, \\
&&\rho={6 \over \ell T}+{6 \over \ell^2}\left(1 - 4 z^2 \right) +
\omega_2(T,z)\,,
\end{eqnarray}
where here and along this paper $\omega_n(T,z)$ will denote functions
that vanish in the limit $T\rightarrow 0$. 
Now, from this density profile we can see that
\begin{eqnarray}
\lim_{T\rightarrow 0} \int_{-T/2}^{T/2}  \rho\;dy = {6 \over \ell},
\end{eqnarray}
as we expected.  Moreover, we can see that the boundary value of the
density satisfies $T\rho|_{\rm T/2} \rightarrow 6/\ell$ in the thin
shell limit, which is the same condition satisfied by the averaged
density, $T \langle \rho \rangle \rightarrow 6/\ell$.

An additional interesting observation for what follows is the following:  
The sets of profiles that yield constant density in the thin-shell limit
(\ref{constant-density-est}) and straight interpolation for the
geometric profile (\ref{straight-interpolation}) coincide. Therefore,
in the thin-shell limit one can assume at the same time a constant
internal structure for the density and a straight-interpolation for
the geometry.

%---------------------------------------------------------------------
\subsection{Einstein-Gauss-Bonnet}
%---------------------------------------------------------------------

Let us move now to the analysis of the same ideas in the presence
of the Gauss-Bonnet term. The Einstein-Gauss-Bonnet 
Lagrangian is 
\begin{eqnarray}
S={1 \over 2 \kappa_5^2 }
\int dx^5 \sqrt{-g} \left[R
-2\Lambda_5 +\alpha \, L_{GB} \right],
\end{eqnarray}
with
\begin{eqnarray}
L_{GB} = R^2 -4R^{AB}R_{AB}+R^{ABCD}R_{ABCD} \,.
\end{eqnarray} 
Now, the field equations deduced from this Lagrangian are 
\begin{eqnarray}
G_{AB}+\alpha H_{AB}=-\Lambda_5 g_{AB}+\kappa_5^2 T_{AB}\,, \label{fieldeq}
\end{eqnarray}
where $H_{AB}$ is the Lanczos tensor \cite{Lanczos}:
\begin{eqnarray}
H_{AB}=2R_{ACDE} R_{B}^{\; CDE}-4R_{ACBD}R^{CD}
-4R_{AC} R_{B}^{\; C}+2RR_{AB}-\frac{1}{2}g_{AB}L_{GB}\, . 
\label{habterm}
\end{eqnarray}
For the ansatz (\ref{brane}) we obtain [compare with equations
(\ref{ttcom}-\ref{yycom}) in Appendix~\ref{appa}]
\begin{eqnarray}
&&3A''(1 - 4\alpha A'^2)-6A'^2(1 - 2\alpha A'^2)=\rho-{6 \over \ell^2},
\label{EGB-equations-rho} \\
&&6A'^2(1 - 2\alpha A'^2)=p_T+{6 \over \ell^2}, \label{EGB-equations-pt} \\
&&p_L=-\rho.
\label{EGB-equations-pl}
\end{eqnarray}
The junction conditions for the geometry are the same as before
(\ref{E-junctions}), implying again the vanishing of the transversal
pressure at the boundaries, $p_T=0$.

In the outside region the solution is a pure adS spacetime
but with a modified length scale
\begin{eqnarray}
{1 \over \tilde\ell} \equiv \sqrt{ {1 \over 4\alpha}
\left(1-\sqrt{1- {8\alpha \over \ell^2}}\right) }.
\end{eqnarray}
Now, isolating $A''$ from~(\ref{EGB-equations-rho}) and~(\ref{EGB-equations-pt})
we can relate the total bending of the geometry on passing through the brane
with the integral of $\rho+p_T$
\begin{eqnarray}
{6 \over \tilde\ell}\left(1- {4 \over 3}{\alpha \over {\tilde\ell}^2}\right)
=(3A'-4 \alpha A'^3)|_{-T/2}^{T/2}=\int_{-T/2}^{T/2}  (\rho+p_T)\; dy.
\end{eqnarray}
Again, if the condition $\rho+p_T \geq 0$ is fulfilled throughout the brane
we will have that in the thin shell limit
\begin{eqnarray}
{6 \over \tilde\ell}\left(1- {4 \over 3}{\alpha \over {\tilde\ell}^2}\right)
=\lim_{T \rightarrow 0} \int_{-T/2}^{T/2}  \rho\; dy.
\label{gb-average-condition}
\end{eqnarray}
At this point we can pursue this analysis in the two simple
cases of constant density profile and straight interpolation.
%

%---------------------------------------------------------------------
\subsubsection{Constant density profile}
%---------------------------------------------------------------------

Following the same steps as before for a constant density profile 
(\ref{constant-density-est})-(\ref{condi}), the equation that one 
has to solve in the thin-shell limit is
\begin{eqnarray}
3A''(1-4\alpha A'^2)=\beta_0(T)-{6 \over \ell^2}+6A'^2(1-2\alpha A'^2)\,.
\end{eqnarray}
Introducing the notation $B\equiv A'$ we reduce this equation to the following
integral
\begin{eqnarray}
y={1 \over 4\alpha } \int_0^B {(4\alpha B^2-1) \; dB
\over B^4 - {1 \over 2\alpha} B^2
- {1 \over 2\alpha} \left({\beta_0(T) \over 6}-{1 \over \ell^2}\right)}\,.
\end{eqnarray}
The result of performing the integration is
\begin{eqnarray}
y={1 \over 2}
\left[
{1 \over \sqrt{-R_{-}}}\tan^{-1}\left({B \over \sqrt{-R_{-}}} \right)
-{1 \over \sqrt{R_{+}}}\tanh^{-1}\left({B \over \sqrt{R_{+}}}\right)
\right],
\end{eqnarray}
where $R_{\pm}$ are
\begin{eqnarray}
R_{\pm}={1 \over 4\alpha }
\left[ 1\pm \sqrt{1+ 8\alpha \left({\beta_0(T) \over 6}-{1 \over \ell^2}\right) }
\right]\,.
\end{eqnarray}
Again, by imposing the boundary condition
\begin{eqnarray}
{T \over 2}={1 \over 2}
\left[
{1 \over \sqrt{-R_{-}}}\tan^{-1}\left({1 \over \tilde\ell \sqrt{-R_{-}}} \right)
-
{1 \over \sqrt{R_{+}}}\tanh^{-1}\left({1 \over \tilde\ell \sqrt{R_{+}}}\right)
\right],
\end{eqnarray}
we find the appropriate form for $\beta_0(T)$. With a lengthy but
straightforward calculation we can check that in the limit $T\rightarrow 0$,
$\beta_0(T)\rightarrow \infty$, we have
\begin{eqnarray}
T\beta_0(T)\rightarrow
{6 \over \tilde\ell}\left(1- {4 \over 3}{\alpha \over {\tilde\ell}^2}\right),
\end{eqnarray}
in agreement with condition (\ref{gb-average-condition}).

Using this same asymptotic expansion, we can see that, in the
thin-shell limit, the profile for $A'(y)$ satisfies
\begin{eqnarray}
A'(y)-{4 \alpha\over 3}A'(y)^3={1 \over 3} \beta_0(T)\; y.
\end{eqnarray}
Recursively, one can create a Taylor expansion for $A'(y)$.
The first two terms are
\begin{eqnarray}
A'(y)={1 \over 3} \beta_0(T)\; y+
{4 \alpha \over 81} \beta_0(T)^3\; y^3+ {\cal O}(y^5)=
{1 \over 3} T \;\beta_0(T)\; z+
{4 \alpha \over 81}T^3 \; \beta_0(T)^3\; z^3 + {\cal O}(z^5).
\end{eqnarray}
By differentiating this expression we find
\begin{eqnarray}
A''(y)={1 \over 3} \beta_0(T)+{4 \alpha \over 27}T^2 \; \beta_0(T)^3\; z^2
+{\cal O}(z^4).
\end{eqnarray}
Now, contrarily to what happens in Einstein theory, this profile
does not correspond to the set considered in the straight interpolation before
(see Fig. 1).
By looking at (\ref{straight-interpolation}) we can identify
\begin{eqnarray}
\gamma_0(T)\equiv{1 \over 3} \beta_0(T), ~~~~
\gamma_1(T)\equiv{4 \alpha \over 27} T^2 \; \beta_0(T)^3.
\end{eqnarray}
Then, we can see that
\begin{eqnarray}
\lim_{T \rightarrow 0} T\gamma_0(T)=
{2 \over \tilde\ell}\left(1- {4 \over 3}{\alpha \over {\tilde\ell}^2}\right)
\neq {2 \over \tilde\ell}, ~~~~
\lim_{T \rightarrow 0} T\gamma_1(T)=
{32 \alpha \over \tilde\ell^3}
\left(1- {4 \over 3}{\alpha \over {\tilde\ell}^2}\right)^3 \neq 0 .
\end{eqnarray}
The coefficients $\gamma_n$ do not satisfy the conditions in
(\ref{straight-interpolation-cond}). Therefore, unlike 
the energy density, the scalar of 
curvature does not have a constant profile.
%====================================================
\begin{figure}[htb]
\vbox{
\hfil
\scalebox{0.7}{\includegraphics{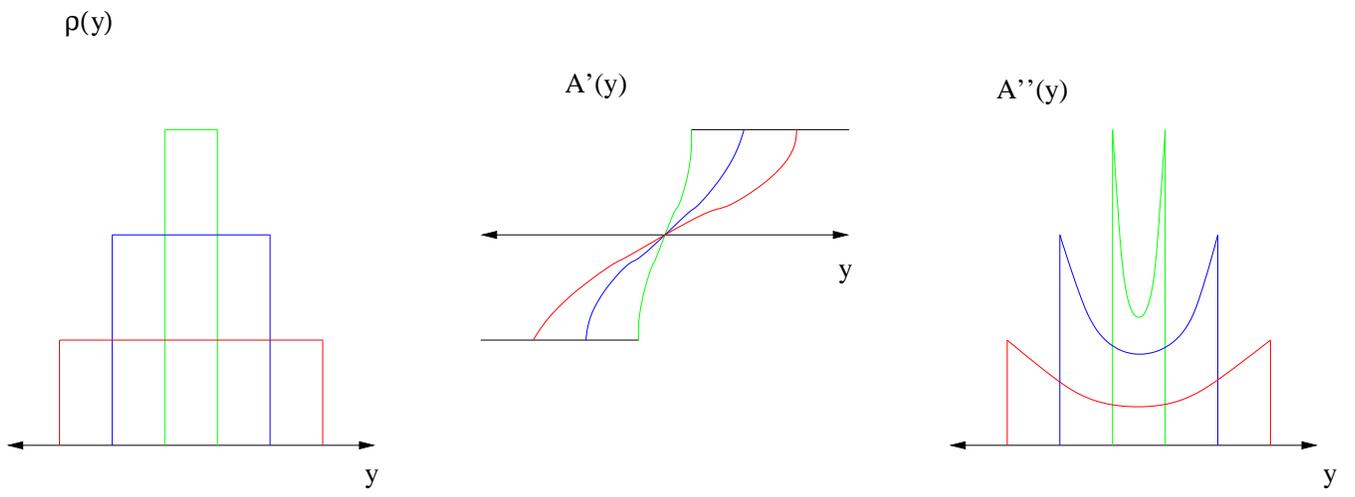}}
\hfil
}
\bigskip
\caption{\label{F:fig1} Family of constant density profiles with decreasing
thickness and associated geometric profile for $A'$ and $A''$.}
\end{figure}
%====================================================

%---------------------------------------------------------------------
\subsubsection{Straight interpolation}
%---------------------------------------------------------------------

As in the Einstein case, the straight interpolation profile for
$A''$ corresponds to
\begin{eqnarray}
A''(z)=\sum_n \gamma_n(T) \; z^{2n},
\label{straight-interpolation2}
\end{eqnarray}
with
\begin{eqnarray}
\lim_{T\rightarrow 0} T ~ \gamma_n(T)\rightarrow 0,
~~~~\forall n\neq 0;~~~~
\lim_{T\rightarrow 0} T ~ \gamma_0(T)\rightarrow {2 \over \tilde\ell}.
\label{straight-interpolation2-cond}
\end{eqnarray}
{}From here we can deduce the associated profiles for $p_T$ and $\rho$
by substituting on (\ref{EGB-equations-rho}) and (\ref{EGB-equations-pt}).

In the limit $T \rightarrow 0$, the dominant part in the density profile
is
\begin{eqnarray}
\rho=3\gamma_0(T)(1-4\alpha \gamma_0(T)^2 \; T^2 \; z^2).
\end{eqnarray}
Identifying
\begin{eqnarray}
\beta_0(T) \equiv 3\gamma_0(T)\,, ~~~~
\beta_1(T) \equiv -12 \alpha  T^2 \; \gamma_0(T)^3\,,
\end{eqnarray}
we find that
\begin{eqnarray}
\lim_{T \rightarrow 0} T \; \beta_0(T) = {6 \over \tilde l}, ~~~~
\lim_{T \rightarrow 0} T \beta_1(T) \neq 0\,.
\end{eqnarray}
Therefore, even in the thin shell limit a straight interpolation
in the geometry does not correspond to a constant density profile (see Fig.2).
In the presence of a Gauss-Bonnet term it is not compatible to
ascribe to have a simple description for the
interior density profile and for the geometric warp factor
at the same time.
In the limit of strictly zero thickness (distributional limit) one will 
unavoidably lose some information on the combined matter-geometry system.
%
%====================================================
\begin{figure}[htb]
\vbox{
\hfil
\scalebox{0.7}{\includegraphics{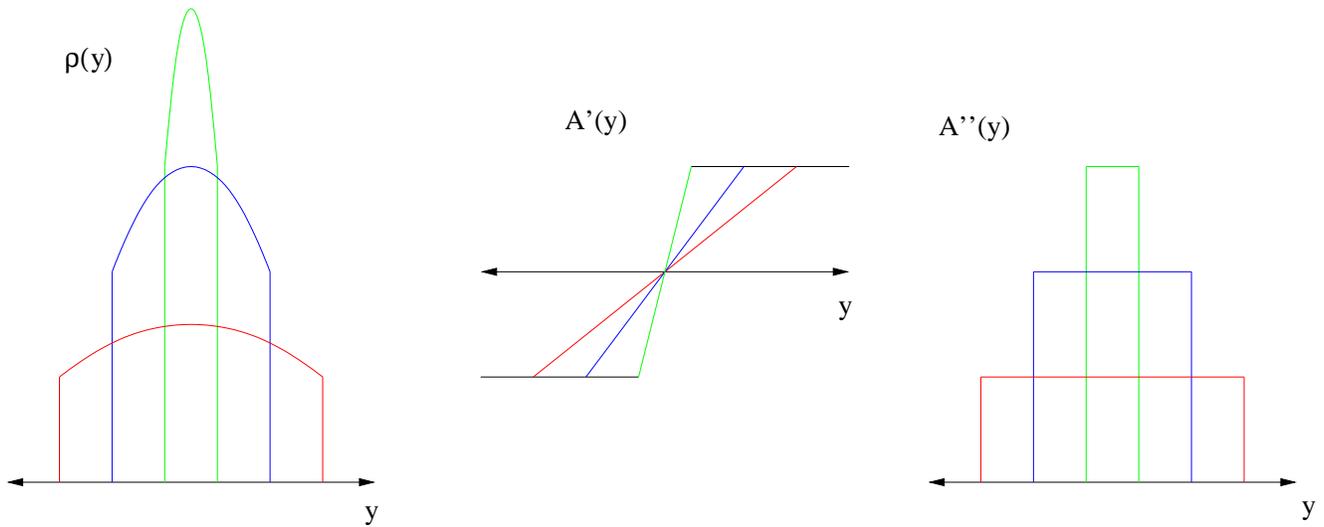}}
\hfil
}
\bigskip
\caption{\label{F:fig2} Plot of the straight interpolation profile
for the geometric factor $A'$ and its associated density profile.}
\end{figure}
%====================================================

To finish this section let us make an additional observation.
{}From expressions (\ref{straight-interpolation2}) and
(\ref{straight-interpolation2-cond}),
we can see that
\begin{eqnarray}
A''= \frac{2}{T} A'|^{}_{T/2} + \nu(T,z) ~~~~\mbox{with}~~~~~
\lim_{T\rightarrow 0} T\nu(T,z) = 0   \,. \label{propsi}
\end{eqnarray}
Using this property in~(\ref{EGB-equations-rho},\ref{EGB-equations-pt}) 
\begin{eqnarray}
\rho +p_T=3A''(1 - 4 \alpha A'^2)={6 \over T} A'(T/2)(1 - 4 \alpha A'^2)
\end{eqnarray}
and evaluating at $y=T/2$ we find
\begin{eqnarray}
T\rho|_{\rm T/2}=6A'(1 - 4 \alpha A'^2)|_{T/2}.
\label{boundary-value}
\end{eqnarray}
We can see that contrarily to what happens in the Einstein case, this condition
is different from the averaged condition (\ref{gb-average-condition})
\begin{eqnarray}
T\langle \rho \rangle=\lim_{T \rightarrow 0} \int_{T/2}^{-T/2}  \rho\; dy=
(3A'-4 \alpha A'^3)|_{-T/2}^{T/2}=
6A'(1-{4 \over 3} \alpha A'^2)\bigg|_{T/2}.
\label{average3}
\end{eqnarray}
Therefore, the averaged density and the boundary density are
different, and this is independent of the brane thickness.  For this simple
model, in thin-shell limit one can define two different internal
density parameters characterizing the thin brane. 
One represents
the total averaged internal density and can be defined as
\begin{eqnarray}
\rho_{\rm av} \equiv \lim_{T \rightarrow 0} T \langle \rho \rangle.
\end{eqnarray}
The other represents an internal density parameter calculated by
extrapolating to the interior the value of the density on the
boundary. This density can be defined as
\begin{eqnarray}
\rho_{\rm bv} \equiv \lim_{T \rightarrow 0} T \rho|_{T/2}.
\end{eqnarray}
The junction conditions for a thin shell given in \cite{Charmousis},
corresponds to the averaged condition (\ref{gb-average-condition}) or
(\ref{average3})
and therefore relate the total bending of the geometry
in passing through the brane with its total averaged density.
Instead, the particular condition analyzed for the boundary value
of $\rho|_{T/2}$ (\ref{boundary-value}) yields in the thin-shell
limit the junction condition in \cite{Germani}. 
This condition is only considering information about the boundary value
of the density and not about its average. 

As a summary, what these analysis suggest is that in the presence
of the Gauss-Bonnet term we cannot forget the interior
structure of the brane, by modelling it by a simple model, even
in the thin-shell limit. We will see again this feature
in the next section on the cosmological dynamics of thick shells.

%---------------------------------------------------------------------
\section{Dynamic thick shells in Einstein and Gauss-Bonnet}
\label{S:dynamic}
%---------------------------------------------------------------------

We are now going to study cosmological thick branes.  To that
end we will use the class of spacetime metrics given in (\ref{themetric}),
which contain a Friedmann-Robertson-Walker (FRW) cosmological model in
every hypersurface $\{y=const.\}$, with a matter content described by
an energy-momentum tensor of the form (\ref{tAB}).  We consider
the additional assumption of a {\em static} fifth dimension: $\dot{b} = 0$.
We can rescale the coordinate $y$ in such a way that $b=1$.
Then, the line element (\ref{themetric}) becomes
\begin{equation}
ds^2=-n^2(t,y)dt^2+a^2(t,y)h_{ij}dx^idx^j+dy^2\,,
\end{equation}
In Appendix~\ref{appa} we show that the $\{ty\}-$component of the
Einstein-Gauss-Bonnet field equations, for the case with a well-defined
limit in Einstein gravity, leads to the equation~(\ref{keyrel}).
In our case it implies the following relation:
\begin{equation}
n(t,y)=\xi(t)\dot a(t,y)\,. \label{0y}
\end{equation}
In this situation the rest of field equations can be written in the form
given in~(\ref{ttcom},\ref{ijcom},\ref{yycom}).  In our case they
become\footnote{The coupling constant $\alpha$ used here is one half the 
one used in \cite{Germani}.}
\begin{eqnarray}
\left[a^4\left(\Phi+2\alpha\Phi^2+\frac{1}{\ell^2}\right)\right]' =
\frac{1}{6}(a^4)'\rho  \,, \label{ttc}
\end{eqnarray}
\begin{eqnarray}
\frac{n'}{n}\frac{\dot{a}}{a'}
\left[a^4\left(\Phi+2\alpha\Phi^2+\frac{1}{\ell^2}\right)\right]'
- {\left[a^4\left(\Phi+2\alpha\Phi^2+\frac{1}{\ell^2}\right)\right]^{\cdot}\,}' =
2\dot{a}a'a^2p_L\,, \label{ijc}
\end{eqnarray}
\begin{eqnarray}
\left[a^4\left(\Phi+2\alpha\Phi^2+\frac{1}{\ell^2}\right)\right]^\cdot =
-\frac{1}{6}(a^4)^\cdot p_T \,, \label{yyc}
\end{eqnarray}
where now $\Phi$ is given by
\begin{eqnarray}
\Phi=\frac{\dot a^2}{n^2 a^2}+\frac{k}{a^2}-\frac{a'^2}{a^2}
= H^2 +\frac{k}{a^2} -\frac{a'^2}{a^2} \,,
\end{eqnarray}
where we have identified the first term with the square of the Hubble function
associated with each $y=const.$ slide,
\begin{eqnarray}
H(t,y)\equiv \frac{\dot a}{n a}\,.
\end{eqnarray}
With the assumption $\dot{b}=0$, the field equation (\ref{ijc}) leads to a
conservation equation for matter of the same form as in the FRW models
[see Eq.(\ref{5ceq})]:
\begin{eqnarray}
\dot{\rho} = -3\frac{\dot{a}}{a}(\rho+p_L) \,. \label{mconeq}
\end{eqnarray}

In the same way we did in the static scenario, we consider here the situation
in which there is a $\mathbb{Z}_2$ symmetry and a fixed proper thickness $T$
for the brane.  Then one has to solve separately the equations for
the {\it bulk}
($|y|>T/2$) and the equations for the thick {\it brane} ($|y|<T/2$).  The first
step has already been done and the result is~\cite{Germani}:
\begin{eqnarray}
\Phi+2\alpha\Phi^2+\frac{1}{\ell^2} = \frac{M}{a^4}\,,~~~~
\mbox{for $|y|>\textstyle{T\over2}$}\,,
\end{eqnarray}
where $M$ is a constant that can be identified with the mass of a black
hole present in the bulk.  Once the solution inside the thick brane has been
found one has to impose the junction conditions~(\ref{cindm},\ref{cincm}) at
$y=\pm T/2$.

The first thing we can deduce from the junction conditions is that the
quantity $\Phi$ is continuous across the two boundaries $y=\pm T/2\,.$
But in general, its transversal derivative, $\Phi'$, will be discontinuous.
Then, using equation (\ref{yyc}) it follows that the transversal pressure
has to be zero on the boundary, $p_T(t,\pm T/2)=0$.  At the same time, from
(\ref{yyc}) we deduce that we must always have:
\begin{eqnarray}
a^4\left(\Phi+2\alpha\Phi^2+\frac{1}{\ell^2}\right)\bigg|_{y=\pm T/2}=M\,.
\label{intm}
\end{eqnarray}
On the other hand, using again the relation~(\ref{keyrel}) we find that
\begin{eqnarray}
H' = -\frac{a'}{a}H~~~~\Rightarrow~~~~\Phi' = -2\frac{a'}{a}
\left(\Phi + \frac{a''}{a}\right)\,,
\end{eqnarray}
and then, expanding (\ref{ttc}), we get
\begin{eqnarray}
(1+4\alpha\Phi)\frac{a''}{a} = \Phi -\frac{2}{\ell^2} -\frac{1}{3}\rho
\,.
\label{mult}
\end{eqnarray}

In the limit $T \rightarrow 0$, the profiles for the density $\rho$ and for 
$a''$ blow up, therefore, these dominant terms in expression~(\ref{mult})
have to be equated. This results in the following equation 
\begin{eqnarray}
(1+4\alpha\Phi)\left({a' \over a}\right)'= -\frac{1}{3}\rho \,.
\label{00-thin-limit}
\end{eqnarray}
In what follows we consider the analysis of the Einstein and
Einstein-Gauss-Bonnet theories separately.

%---------------------------------------------------------------------
\subsection{Einstein}
%---------------------------------------------------------------------

In Einstein gravity it is not difficult to write down an equation describing
the dynamics of every layer in the interior of a thick shell. 
To that end we will take $\alpha=0$ in the equations above.
By integrating (\ref{ttc}) over the interval $(-T/2,y_\ast)$
and using (\ref{intm},\ref{00-thin-limit}) we arrive at
\begin{eqnarray}
\left(H^2 +{k \over a^2} + {1 \over \ell^2}  \right)=
\left({a' \over a}\right)^2+
{M \over a^4}+{1 \over 6 a^4}\int_{-T/2}^{y_\ast} (a^4)'\rho\, dy=
{1 \over 36} \left(\int_{-y_\ast}^{y_\ast} \rho\, dy \right)^{2}+
{M \over a^4}+{1 \over 6a^4}\int_{-T/2}^{y_\ast} (a^4)'\rho\, dy\,.
\label{layer-Einstein}
\end{eqnarray}
A particular layer of matter inside the shell, located at $y=y_\ast$,
can be seen as separating an internal spacetime from a piece of
external spacetime.  From the previous equation we can see that the
cosmological evolution of each layer $y=y_\ast$ in the thick shell
depends on the balance between the integrated density beyond the layer
(external spacetime) and a weighted contribution of the integrated
density in the internal spacetime. Therefore, the dynamics of each
shell layer will be influenced by the particular characteristics of
the internal density profile inside the shell. However, by looking at
this same equation we can see that the dynamics of the boundary layer
$y_\ast=-T/2$ is only influenced by the total integrated density
throughout the shell:
\begin{eqnarray}
\left(H^2 +{k \over a^2} + {1 \over \ell^2}- 
{M \over a^4}  \right)\bigg|_{-T/2}=
{1 \over 36}\left(\int_{-T/2}^{T/2} \rho dy\right)^{2}=
{1 \over 36}\left(T\langle \rho \rangle \right)^{2}={1 \over 36}\rho_{\rm av}^2
\end{eqnarray}
This is the modified Friedmann equation for the cosmological evolution
of the brane \cite{Binetruy}.

%---------------------------------------------------------------------
%\subsubsection{Constant density profile}
%---------------------------------------------------------------------

In the same manner as we proceeded with static shells, let us analyze
the case in which the density profile tends to a (time-dependent) constant
in the thin-shell limit:
\begin{eqnarray}
\rho =\beta_0(T,t)+ \omega_3(T,t,z), ~~~~
\lim_{T \rightarrow 0} T \beta_0(T,t)=\rho_{\rm av}(t)\,, ~~~~
\lim_{T \rightarrow 0} T\omega_3(T,t,z) =0\,.
\label{constant-density}
\end{eqnarray}
When $\alpha=0$, equation~(\ref{00-thin-limit}) tells us that if the
density profile depends only on $t$ in the thin-shell limit, then, in
this same limit, the blowing up part of the geometry $(a'/a)'$ is also
constant through the brane interior, describing what we called before a
straight interpolation. A simple density profile amounts to a simple
and equivalent geometric profile and vice-versa. In this same case but
including the Gauss-Bonnet term, $\alpha \neq 0$, the geometrical factor $(a'/a)'$
will exhibit a non trivial profile in $y$, even in the thin-shell limit.
We will see this fact in more detail in the next subsection.

Now, in the case in which $\rho$ depends only on time,
equation~(\ref{layer-Einstein}) reads
\begin{eqnarray}
\left.\left(H^2 +{k \over a^2} + {1 \over \ell^2}  \right)\right|_{y_\ast}=
\left.\left({a' \over a}\right)^2\right|_{y_\ast}+ {M \over a^4(y_\ast)}+
\frac{1}{6}\rho(t)\left.\left(1+ {a^4_{T/2} \over a^4}\right)\right|_{y_\ast}\,.
\end{eqnarray}
{}From equation~(\ref{00-thin-limit}) we deduce that
\begin{eqnarray}
\left.{a' \over a}\right|_{y_\ast}={1 \over 6}\rho_{\rm av} -
{1 \over 3}\int_{-T/2}^{y_\ast} \rho \, dy
=- {1 \over 3}\rho_{\rm av} z_\ast\,,
\end{eqnarray}
and integrating we obtain
\begin{eqnarray}
a(t,y)=a_0(t)\exp\left(-{1 \over 6}\rho_{\rm av}(t)Tz^{2}  \right),
\end{eqnarray}
(remember that $z\equiv y/T$).
Therefore, in the lowest order in $T$ we have an equation for the
internal geometry of the following form
\begin{eqnarray}
H_0^2+{k \over a_0^2}+{1 \over \ell^2}={1 \over 9}\rho_{\rm av}^2 z^2 +
{1 \over 36}\rho_{\rm av}^2(1-4z^2)+\frac{M}{a_0^4}
={1 \over 36}\rho_{\rm av}^2+\frac{M}{a_0^4},
\end{eqnarray}
which is exactly the standard braneworld generalized Friedmann equation~\cite{Binetruy}.

%---------------------------------------------------------------------
\subsection{Einstein-Gauss-Bonnet}
%---------------------------------------------------------------------

In the general Einstein-Gauss-Bonnet case, equation~(\ref{00-thin-limit}) can be
written as
\begin{eqnarray}
\left[1+4\alpha\left(H^2+{k \over a^2}\right)\right]\left({a' \over a}\right)'
-4\alpha \left({a' \over a}\right)^2 \left({a' \over a}\right)'
=-{1 \over 3}\rho.
\label{gb-expanded}
\end{eqnarray}
Then, integrating between $-T/2$ and $T/2$ yields
\begin{eqnarray}
2\left[1+4\alpha\left(H^2+{k \over a^2}\right)\right]
\left({a' \over a}\right)\bigg|_{T/2}
-{8\alpha \over 3}\left({a' \over a}\right)^3 \bigg|_{T/2}
=-{1 \over 3}\langle \rho \rangle T=-{1 \over 3}\rho_{\rm av}\,.
\label{gb-blowup}
\end{eqnarray}
The boundary equation (\ref{intm}) can be written as
\begin{eqnarray} 
\left.\left(H^2+{k \over a^2}\right)\right|_{T/2}-
\left.\left({a' \over a}\right)^2\right|_{T/2}
+2\alpha\left[\left(H^2+{k \over a^2}\right)-\left({a' \over a}\right)^2
\right]^2_{T/2}-{M \over a^4(T/2)}+{1 \over \ell^2}=0\,, \label{bcegb}
\end{eqnarray}
This is a quadratic equation for $(a'/a)^2|_{T/2}$ with solutions
\begin{eqnarray}
\left({a' \over a}\right)^2\bigg|_{T/2}={1 \over 4 \alpha}
\left.\left[1+4\alpha\left(H^2+{k \over a^2}\right)\right|_{T/2}
\pm
\sqrt{1+{8\alpha \over \ell^2}-{8\alpha M \over a^4}} \right]\,.
\end{eqnarray}
{}From these two roots we will take only the minus sign as it is the only
one with a well defined limit when $\alpha$ tends to zero.
Now, by squaring (\ref{gb-blowup}) and substituting the previous
root we arrive to a cubic equation for $H^2+k/a^2$ first found
in~\cite{Charmousis}. This cubic equation has a real root
that can be expressed as~\cite{Gregory-Padilla} 
\begin{eqnarray}
H^2+{k \over a^2}={1 \over 8 \alpha}
\left[
(\sqrt{\lambda^2+\zeta^3}+\lambda)^{2/3}
+
(\sqrt{\lambda^2+\zeta^3}-\lambda)^{2/3}
-2
\right]\,, \label{feqchar}
\end{eqnarray}
where
\begin{eqnarray}
\lambda \equiv \sqrt{\alpha \over 2} \rho_{\rm av}\,,
~~~~
\zeta \equiv  \sqrt{1+8\alpha V(a) } \equiv
\sqrt{1+{8\alpha \over \ell^2}-{8\alpha M \over a^4}}\,.
\end{eqnarray}
In addition to this equation we need the conservation equation
\begin{eqnarray}
\dot \rho = 3 H (\rho+p_L)\,,
\end{eqnarray}
which is valid for each section $y=y_*$, and in particular, 
for the boundary, $y=T/2$.
This equation can be averaged to give
\begin{eqnarray}
T\langle \dot \rho \rangle = 3 \langle H T (\rho+p_L) \rangle = 3 H|_{T/2}
(T \langle \rho \rangle+T \langle p_L \rangle)+{\cal O}(T),
\end{eqnarray}
or written in another way
\begin{eqnarray}
\dot \rho_{\rm av}= 3 H|_{T/2}(\rho_{\rm av}+p_{{\rm av}L})\,.
\end{eqnarray}
This happens because
\begin{eqnarray}
H(t,y)\rightarrow H_0(t,y_0)+{\cal O}(T)
\end{eqnarray}
for whatever $y_0 \in[-T/2,T/2]$, which we have taken as $y_0=T/2$ for convenience.

Let us analyze now the simple case of a constant density profile
(\ref{constant-density}). For consistency with the $T \rightarrow 0$
case, we know that
\begin{eqnarray}
a(t,y)=a_0(t)[1+T{\tilde a}(t,z)]+ {\cal O}(T^2)\,,
\end{eqnarray}
and therefore, from~(\ref{0y})
\begin{eqnarray}
n(t,y)= \xi(t)\left[a_0\left(1+T\tilde{a}(t,z)\right)\right]^\cdot\,.
\end{eqnarray}
Then, in the same limit equation~(\ref{00-thin-limit}) reads
\begin{eqnarray}
{\tilde a}_{,zz}= -{1\over 3}
{\beta_0(T,t)T  \over \left[1+4\alpha
\left(H_0^2+{k \over a_0^2}-{\tilde a}_{,z}^2 \right)\right]}.
\end{eqnarray}
(Here the subscript $,z$ denotes differentiation with respect to $z$.)
A necessary condition to have a straight interpolation for the geometry
is that ${\tilde a}(t,z)=b(t)Z(z)$. To check whether or not a simple density
profile corresponds to a simple geometrical profile we can therefore try
to solve the previous equation by separation of variables.
It is not difficult to see that in order to find a solution
with a well defined Einstein limit we need that
\begin{eqnarray}
b(t) = \mu \,,
~~~~ \mu^{-1}\beta_0(T,t)T = \mu^{-1}\beta_0(T)T = \rho_{\rm av}: {\rm constant},
~~~~ H_0^2+{k \over a_0^2}=\Lambda_4\,,
\end{eqnarray}
where $\mu$ is a constant that can be absorbed into the function $Z(z)$,
so we will take it to be $\mu=1$. 
In this way we will recover the anti-de Sitter and de Sitter solutions
for the brane (depending on the sign of the effective four-dimensional
cosmological constant).  To find the specific $y$ profile we have to solve
\begin{eqnarray}
Z_{,zz}= -{1\over 3}
{\rho_{\rm av}  \over \left[1+4\alpha
\left(\Lambda_4-Z_{,z}^2\right)\right]}.
\end{eqnarray}
This equation can be integrated to get
\begin{eqnarray}
(1+4\alpha\Lambda_4)Z_{,z}-{4\alpha \over 3}Z_{,z}^3=-{1\over 3}\rho_{\rm av}\, z\,.
\end{eqnarray}
For our proposes the specific solution of this cubic equation is
not important.  What we want to point out is that the solution does not
correspond to a straight interpolation as it happened in the Einstein case.
So, in general, simple solutions for the matter profile lead to non trivial
profiles for the scalar curvature even in the thin-shell limit.

Let us see now what happens when we take a simple model for the
geometry, the straight interpolation model:
\begin{equation}
a(t,z)=a_0(t)-\frac{1}{2}b(t)z^2T.
\label{req}
\end{equation}
{}From equation~(\ref{gb-expanded}) we can deduce the density profile
in this situation
\begin{eqnarray}
\lim_{T \rightarrow 0}T \rho=
3\left[1+4\alpha\left(H_0^2+{k \over a_0^2}\right)\right]
\left({b \over a_0}\right)-12\alpha\left({b \over a_0}\right)^3 z^2.
\end{eqnarray}
As in the static case, even for very small thickness the density
profile has now a non-trivial structure.
We can observe that
\begin{eqnarray}
{a' \over a}\bigg|_{T/2}= -{1 \over 2}{b \over a_0} + {\cal O}(T) \,,
~~~~
\left({a' \over a}\right)'= {2 \over T}\left[ {a' \over a}\bigg|_{T/2}
+{\cal O}(T)\right]\,. \label{extraas}
\end{eqnarray}
The second relation and equation~(\ref{propsi}) coincide in the thin-shell
limit. Therefore, evaluating~(\ref{gb-expanded}) on $y=T/2$ we obtain
\begin{eqnarray}
\left[1+4\alpha\left(H^2+{k \over a^2}\right)\right]\left.\left({a' \over a}\right)
\right|_{T/2}
-4\alpha \left.\left({a' \over a}\right)^3\right|_{T/2}
=-{1 \over 6}T \rho|_{T/2}=-{1 \over 6}\rho_{\rm bv}\,. \label{cubic2}
\end{eqnarray}
Now, following the same steps that we followed previously but using this condition
instead of (\ref{gb-blowup}) we arrive at a cosmological generalized
Friedmann equation \cite{Germani} different from that in~\cite{Charmousis} in its form
and in the fact that it depends on the quantity associated with the boundary value of
the energy density, $\rho_{\rm bv}$, instead of the value associated with the
average of the energy density, $\rho_{\rm av}$.  Remarkably, the cubic equation
that results from combining the last equation~(\ref{cubic2}) with the boundary
condition~(\ref{bcegb}) becomes in this case linear.  That is, the coefficients of the 
terms quadratic and cubic in $H^2+k/a^2$ vanish~\cite{Germani}.  The modified 
Friedmann equation found in this case is:
\begin{eqnarray}
H^2+{k \over a^2}=
{1 \over \left(1+{8\alpha \over \ell^2}-{8\alpha M \over a^4}\right)}
{1 \over 36} \rho_{\rm bv}^2 +{1 \over 4 \alpha}\left(
\sqrt{1+{8\alpha \over \ell^2}-{8\alpha M \over a^4}}-1 \right)\,.
\end{eqnarray}
In contrast with the modified Friedmann equation~(\ref{feqchar}), which
was obtained by using a completely general procedure, in order to obtain
this equation we had to use a procedure which required to consider an
extra assumption, namely equation~(\ref{extraas}), and hence, it will
not work  for profiles of the metric function $a(t,z)$ that do not 
satisfy these requirements or equivalent ones.   On the other hand,
by looking at the developments here presented, we can conclude that the 
different results found in the literature
for the dynamics of a distributional shell have their origin in the
additional internal richness introduced in the brane by the presence
of the GB term.

%----------------------------------------------------------------
\section{Summary}
\label{S:summary}
%----------------------------------------------------------------

%
We have analyzed and compared how the thin shell limit of static and
cosmological braneworld models is attained in Einstein and
Einstein-Gauss-Bonnet gravitational theories. We have seen that the
generalized Friedmann equation proposed in~\cite{Charmousis} is always
valid and relates the dynamical behaviour of the shell's boundary with
its total internal density (obtained by integrating transversally the
density profile).  Instead, the generalized Friedmann equation
proposed in~\cite{Germani} relates the dynamical behaviour of the
shell's boundary with the boundary value of the density within the
brane. This equation is not always valid, only for specific
geometrical configurations.

Einstein equations in these models transfer the blowing up
contributions of the thin-shell internal density profile to the
structure of the internal geometry in a faithful way. If we don't know
the internal structure of the shell we can always model it in simple
terms by assuming an (almost) constant density profile and an (almost)
constant internal curvature. However, the GB term makes incompatible
to have both magnitudes (almost) constant.  If the density is (almost)
constant, then the curvature is not, and vice-versa. Therefore, we
can say that the particular structure of the Einstein-Gauss-Bonnet theory 
introduces important microphysical features on the matter-geometry
configurations beyond those in Einstein gravity, that are hidden in the
distributional limit.

%----------------------------------------------------------------
\section*{Acknowledgments}
%----------------------------------------------------------------

The authors wish to thank Roy Maartens for useful discussions and
comments.  CB is supported by the EC under contract
HPMF-CT-2001-01203.  CG is supported by the PPARC and wishes to thank
C\'edric Deffayet, Nathalie Deruelle and Alfredo Germani for
enlightening conversations on the subject of this paper, and to the
Institut d'Astrophysique de Paris (IAP) and the Dipartimento di Ingegneria
Aereospaziale e Astronautica, Universit\'a ``La Sapienza'' di Roma, 
for hospitality in visits during the realization of this work.
CFS is supported by the EPSRC.

%----------------------------------------------------------------
%---------------------------  APPENDICES   ----------------------

\appendix

\section{The 5D metric and its associated geometrical quantities and field equations}\label{appa}

In this appendix we present the main geometrical quantities and field
equations associated with the following 5D metric
\begin{eqnarray}
ds^2 = g_{AB}dx^Adx^B = -n^2(t,y)dt^2 + a^2(t,y)h_{ij}(x^k)dx^idx^j + b^2(t,y)dy^2\,,
\label{themetric}
\end{eqnarray}
where $h_{ij}$ is the metric of the three-dimensional maximally symmetric
surfaces $\{t=const.,y=const.\}$, whose spatial curvature is parametrized
by $k=-1,0,1$.  A particular representation of $h_{ij}$ is
\begin{eqnarray}
h_{ij}dx^idx^j = \frac{1}{\left(1+\frac{k}{4}r^2\right)^2}\left(dr^2 +
r^2d\Omega^2_2\right)\,,
\end{eqnarray}
being $d\Omega^2_2$ the metric of the 2-sphere.  The metric~(\ref{themetric})
contains as particular cases the metrics used along this paper.

The non-zero components of the Einstein tensor $G_{AB}$ corresponding to this line
element are given by ($\dot{Q}=\partial_tQ$, $Q'=\partial_yQ$):
\begin{eqnarray}
G_{tt} & = & 3\left\{ n^2\Phi + \frac{\dot{a}}{a}\frac{\dot{b}}{b}-
\frac{n^2}{b^2}\left[\frac{a''}{a}-\frac{a'}{a}\frac{b'}{b}\right]\right\}
\,, \nonumber \\
G_{ty} & = & 3\left( \frac{\dot{a}}{a}\frac{n'}{n}+\frac{a'}{a}\frac{\dot{b}}{b}
-\frac{\dot{a}'}{a}\right) \,, \nonumber \\
G_{ij} & = & \frac{a^2}{b^2}h_{ij}\left\{\frac{a'}{a}\left(\frac{a'}{a}
+2\frac{n'}{n}\right)-\frac{b'}{b}\left(\frac{n'}{n}+2\frac{a'}{a}\right)
+2\frac{a''}{a} + \frac{n''}{n}\right\}
\nonumber \\
& - & \frac{a^2}{n^2}h_{ij}\left\{\frac{\dot{a}}{a}\left(\frac{\dot{a}}{a}-
2\frac{\dot{n}}{n}\right) - \frac{\dot{b}}{b}\left(\frac{\dot{n}}{n}
-2\frac{\dot{a}}{a}\right) + 2\frac{\ddot{a}}{a}+\frac{\ddot{b}}{b} \right\}
\nonumber \\
& - & kh_{ij}\,, \nonumber \\
G_{yy} & = & 3\left\{-b^2\Phi + \frac{a'}{a}\frac{n'}{n} - \frac{b^2}{n^2}
\left[\frac{\ddot{a}}{a}-\frac{\dot{a}}{a}\frac{\dot{n}}{n}\right]\right\} \,,
\end{eqnarray}
where
\begin{equation}
\Phi(t,y) = \frac{1}{n^2}\frac{\dot{a}^2}{a^2}-\frac{1}{b^2}\frac{a'^2}{a^2}
+\frac{k}{a^2}\,. \label{defphi}
\end{equation}
Apart from the metric and the Einstein tensor, the field equations in
Einstein-Gauss-Bonnet gravity~(\ref{fieldeq}) contain a term quadratic in the
curvature, namely $H_{AB}$ [see Eq.~(\ref{habterm})].
The non-zero components of this tensor can be written as follows
\begin{eqnarray}
H_{tt} & = & 6\Phi\left[
\frac{\dot{a}}{a}\frac{\dot{b}}{b}+\frac{n^2}{b^2}\left(
\frac{a'}{a}\frac{b'}{b}-\frac{a''}{a} \right) \right]\,, \nonumber \\
H_{ty} & = &
6\Phi\left(\frac{\dot{a}}{a}\frac{n'}{n}+\frac{a'}{a}\frac{\dot{b}}{b}
-\frac{\dot{a}'}{a}\right)\,, \nonumber \\
H_{ij} & = & 2a^2h_{ij}\left\{ \Phi\left[\frac{1}{n^2}\left(\frac{\dot{n}}{n}
\frac{\dot{b}}{b} - \frac{\ddot{b}}{b}\right) - \frac{1}{b^2}\left(
\frac{n'}{n}\frac{b'}{b}-\frac{n''}{n}\right)\right] \nonumber \right. \\
& + & \frac{2}{a^2bn}\left[ \frac{\dot{a}^2\dot{b}\dot{n}}{n^4}
+\frac{a'^2b'n'}{b^4} + \frac{\dot{a}a'}{b^2n^2}\left(
b'\dot{n}-\dot{b}n'\right) \right] \nonumber \\
& - &  2\left[ \frac{1}{n^2}\frac{\ddot{a}}{a}\left(
\frac{1}{n^2}\frac{\dot{a}}{a}
\frac{\dot{b}}{b} + \frac{1}{b^2}\frac{a'}{a}\frac{b'}{b}\right)-\frac{1}{b^2}
\frac{a''}{a}\left(\frac{1}{n^2}\frac{\dot{a}}{a}\frac{\dot{n}}{n}
+\frac{1}{b^2}\frac{a'}{a}\frac{n'}{n}\right)\right] \nonumber \\
& + & \left. \frac{2}{b^2n^2}\left[\frac{\ddot{a}}{a}\frac{a''}{a}-
\frac{\dot{a}^2}{a^2}\frac{n'^2}{n^2}-\frac{a'^2}{a^2}\frac{\dot{b}^2}{b^2}
-\frac{\dot{a}'}{a}\left(\frac{\dot{a}'}{a}-2\frac{\dot{a}}{a}\frac{n'}{n}
-2\frac{a'}{a}\frac{\dot{b}}{b}\right) \right] \right\}\,, \nonumber \\
H_{yy} & = & 6\Phi\left[\frac{a'}{a}\frac{n'}{n}+\frac{b^2}{n^2}\left(
\frac{\dot{a}}{a}\frac{\dot{n}}{n}-\frac{\ddot{a}}{a}\right)\right]\,.
\end{eqnarray}

In this paper we consider the situation in which a thick brane is embedded
in the five-dimensional spacetime described by (\ref{themetric}), whose boundaries
are located at $y=const.$ hypersurfaces.  Let us consider the usual junction
conditions at a hypersurface $\Sigma_{y_c}\equiv\{p\in V_5 \,|\, y(p)=y_c\}$,
that is, the continuity of the induced metric, $q_{AB}=g_{AB}-n_An_B$ and the
extrinsic curvature, $K_{AB} = -q^{C}_{(A}q^D_{B)}\nabla^{}_C n^{}_D$,
of $\Sigma_{y_c}$:
\begin{eqnarray}
n(t,y^+_c) = n(t,y^-_c) \,, ~~~~ a(t,y^+_c) = a(t,y^-_c) \,, 
\label{cindm}
\end{eqnarray}
\begin{eqnarray}
\frac{n'(t,y^+_c)}{b(t,y^+_c)} = \frac{n'(t,y^-_c)}{b(t,y^-_c)} \,, ~~~~
\frac{a'(t,y^+_c)}{b(t,y^+_c)} = \frac{a'(t,y^-_c)}{b(t,y^-_c)} \,. 
\label{cincm}
\end{eqnarray}

Now, let us assume a matter content described by an energy-momentum tensor
of the form
\begin{eqnarray}
\kappa^2_5\, T_{AB} = \rho u_{A}u_{B} + p_Lh_{AB} + p_Tn_{A}n_{B}\,,
\end{eqnarray}
where
\begin{eqnarray}
u_A=(-n(t,y),\mb{0},0)\,,~~~~
h_{AB} = g_{AB} + u_Au_B - n_An_B \,,~~~~
n_A=(0,\mb{0},b(t,y))\,,
\end{eqnarray}
where $\rho\,,$ $p_L\,,$ and $p_T$ denote, respectively, the energy density and
the longitudinal and transverse pressures with respect to the
observers $u^A$.  They are functions of $t$ and $y$.  The energy-momentum 
conservation equations, $\nabla_A T^{AB}=0$, reduce to the following two equations:
\begin{eqnarray}
\dot{\rho} = -\frac{\dot{b}}{b}(\rho+p_T) -3\frac{\dot{a}}{a}(\rho+p_L)\,,
\label{5ceq}
\end{eqnarray}
\begin{eqnarray}
p_T' = -3\frac{a'}{a}(p_T-p_L)-\frac{n'}{n}(\rho+p_T) \,.
\end{eqnarray}

In this situation, the $\{ty\}$-component of the field equations for the metric
(\ref{themetric}) in Einstein-Gauss-Bonnet gravity [Eq.~(\ref{fieldeq})] has the
form
\begin{eqnarray}
\left(1+4\alpha\Phi\right)\left(\frac{\dot{a}}{a}\frac{n'}{n}+
\frac{a'}{a}\frac{\dot{b}}{b}-\frac{\dot{a}'}{a}\right) = 0\,. \label{tycom}
\end{eqnarray}
If we discard the possibility $1+4\alpha\Phi=0$ by restricting ourselves
to models with a well-defined limit in Einstein gravity ($\alpha\rightarrow
0$), we have that the metric functions must satisfy the following
relation
\begin{eqnarray}
\dot{a}' = \frac{n'}{n}\dot{a} + \frac{\dot{b}}{b}a' \,. \label{keyrel}
\end{eqnarray}
Using this consequence of the $\{ty\}$-component, we can rewrite the
rest of components of $G_{AB}$ and $H_{AB}$ as
\begin{eqnarray}
G_{tt} = \frac{3n^2}{2a^3a'}\left(a^4\Phi\right)'\,,~~~~
G_{yy} = - \frac{3b^2}{2a^3\dot{a}}\left(a^4\Phi\right)^\cdot   \,,
\end{eqnarray}
\begin{eqnarray}
G_{ij} = \frac{1}{2\dot{a}a'}h_{ij}\left\{\frac{\dot{b}}{b}\frac{a'}{\dot{a}}
\left(a^4\Phi\right)^\cdot + \frac{n'}{n}\frac{\dot{a}}{a'}
\left(a^4\Phi\right)' - \left(a^4\Phi\right)^\cdot{}'  \right\}   \,,
\end{eqnarray}
\begin{eqnarray}
H_{tt} = \frac{3n^2}{2a^3a'}\left(a^4\Phi^2\right)'\,,~~~~
H_{yy} = - \frac{3b^2}{2a^3\dot{a}}\left(a^4\Phi^2\right)^\cdot   \,,
\end{eqnarray}
\begin{eqnarray}
H_{ij} = \frac{1}{2\dot{a}a'}h_{ij}\left\{\frac{\dot{b}}{b}\frac{a'}{\dot{a}}
\left(a^4\Phi^2\right)^\cdot + \frac{n'}{n}\frac{\dot{a}}{a'}
\left(a^4\Phi^2\right)' - \left(a^4\Phi^2\right)^\cdot{}'  \right\}   \,.
\end{eqnarray}

Then, the field equations~(\ref{fieldeq}) for the metric~(\ref{themetric})
are equivalent to equation (\ref{keyrel}) and the following three equations
\begin{eqnarray}
\left[a^4\left(\Phi+2\alpha\Phi^2+\frac{1}{\ell^2}\right)\right]' =
\frac{1}{6}(a^4)'\rho  \,, \label{ttcom}
\end{eqnarray}
\begin{eqnarray}
\frac{\dot{b}}{b}\frac{a'}{\dot{a}}
\left[a^4\left(\Phi+2\alpha\Phi^2+\frac{1}{\ell^2}\right)\right]^\cdot
+ \frac{n'}{n}\frac{\dot{a}}{a'}
\left[a^4\left(\Phi+2\alpha\Phi^2+\frac{1}{\ell^2}\right)\right]'
- {\left[a^4\left(\Phi+2\alpha\Phi^2+\frac{1}{\ell^2}\right)\right]^\cdot{}}' =
2\dot{a}a'a^2p_L\,, \label{ijcom}
\end{eqnarray}
\begin{eqnarray}
\left[a^4\left(\Phi+2\alpha\Phi^2+\frac{1}{\ell^2}\right)\right]^\cdot =
-\frac{1}{6}(a^4)^\cdot p_T \,. \label{yycom}
\end{eqnarray}
Introducing (\ref{ttcom}) and (\ref{yycom}) into (\ref{ijcom}) we get
the conservation equation~(\ref{5ceq}).

%========================================================================
% When possible, the references have proper Spires citations attached.
% This is supposed to help the Spires staff in updating their database.
% Don't touch the commented ``citation = '' lines.
%========================================================================

%----------------------------------------------------------------

%----------------------------------------------------------------
\end{document}